\documentclass[aps, prl, twocolumn, superscriptaddress, nofootinbib, amsmath, amssymb]{revtex4-1}

\usepackage{graphicx}  
\usepackage{mathrsfs} 
\usepackage[normalem]{ulem} 
\usepackage{bigints}
\usepackage{amsmath,amssymb}
\usepackage{mathtools}
\usepackage{bbm}
\newcommand{\beq}{\begin{eqnarray}}
\newcommand{\eeq}{\end{eqnarray}}

\begin{document}
	
	\title{Spectral Methods and Running Scales in Causal Dynamical Triangulations}
	
	\author{Giuseppe Clemente}
	\email{giuseppe.clemente@pi.infn.it}
	\affiliation{Dipartimento di Fisica dell'Universit\`a di Pisa and INFN
		- Sezione di Pisa, Largo Pontecorvo 3, I-56127 Pisa, Italy.}
	
	\author{Massimo D'Elia}
	\email{massimo.delia@unipi.it}
	\affiliation{Dipartimento di Fisica dell'Universit\`a di Pisa and INFN
		- Sezione di Pisa, Largo Pontecorvo 3, I-56127 Pisa, Italy.}
	
	\author{Alessandro Ferraro}
	\email{alessandro.ferraro@pi.infn.it}
	\affiliation{Dipartimento di Fisica dell'Universit\`a di Pisa and INFN
		- Sezione di Pisa, Largo Pontecorvo 3, I-56127 Pisa, Italy.}

	\begin{abstract}
		The spectrum of the Laplace-Beltrami operator,
		computed on the spatial slices of 
		Causal 
		Dynamical Triangulations, is a powerful probe of 
		the geometrical properties of the configurations
		sampled in the various phases of the lattice theory.
		We
		study the behavior of the lowest eigenvalues
		of the spectrum and show that this can provide information
		about the running of length scales
		as a function of the bare parameters of the theory,
		hence about 
		the critical behavior around possible second order transition points 
		in the CDT phase diagram, where a continuum limit could be defined.
	\end{abstract}

	\maketitle
	
	\emph{Introduction} - The quest for a self-consistent Quantum Theory
	of Gravity is still far from being settled. Field theoretical
	approaches are not able to provide a theory which is renormalizable 
	from a standard perturbative point of view, 
	i.e.~where all ultraviolet (UV) divergences 
	are reabsorbed at all orders in the coupling expansion
	by the addition of a finite number of counterterms~\cite{sagnotti}.
	However, the idea that a non-perturbative solution could still
	be found within Quantum Field Theory has not yet been 
	given up, a promising approach being represented 
	by the so-called {\em asymptotic safety} program~\cite{ass_weinberg}.
	
	The program is rooted in the renormalization group (RG) framework:
	in a few words, the idea of asymptotic safety is to 
	find a non-perturbative UV fixed point in the space of possible 
	parameters, with an RG-flow line stemming from it and reproducing
	the theory of gravity at lower energy scales.
	Consistent progress has been achieved along this direction
	by approaches studying the RG-flow~\cite{ass_Litim, frg_Reuter98, frg_qeg, frg_Bonanno02}.
	
	A different and complementary approach to the above program 
	is numerical: one considers a discretization of the 
	Euclidean path integral of the theory in configuration space,
	suitable to be investigated by Monte-Carlo simulations,
	and looks for possible critical points, i.e.~for 
	values of the bare parameters where the correlation length,
	measured in units of the elementary discretization
	scale, diverges. Such points are candidate UV fixed points
	where a continuum limit for QG could be taken.
	
	A standard discretization is based
	on the Regge formalism~\cite{regge}: space-time configurations (triangulations)
	are represented by the possible collections of flat simplexes,
  glued together so as to 
	reproduce different possible geometries.
	Pioneering work in this direction has been done by the 
	approach called Dynamical Triangulations~\cite{edt1,edt2,edt3,dt_forcrand,dt_syracuse}.
	In the particular approach known as
	Causal Dynamical Triangulations (CDT)~\cite{cdt_report,cdt_pioneer,cdt_secondord,cdt_secondfirst,cdt_newhightrans,cdt_charnewphase,cdt_toroidal}, 
	the causal condition of global hyperbolicity~\cite{causconds,cdt_nofoliae} is additionally enforced 
	on triangulations by means of a space-time foliation,
	with spatial slices characterized by a fixed
	topology (usually $S^3$), and typically periodic boundary conditions (p.b.c.)
	in the time direction. 
	
	Such configurations
	are then sampled according to  
	a discretized version of the candidate
	continuum action.
	In absence of matter fields, 
	the simplest candidate is the Einstein-Hilbert action,
	whose discretized version reads
	\begin{equation}\label{eq:4Daction}
	S_E = -k_0 N_0 + k_4 N_4 + \Delta ( N_4 + N_{41}-6 N_0) \, ,
	\end{equation}
	where $N_0$, $N_4$ and $N_{41}$ count respectively 
	the total number of vertices, of generic pentachorons and 
	of special pentachorons having four vertices on the same {spatial slice}, while
	$k_4$, $k_0$ and $\Delta$ are free dimensionless parameters, 
  related to the {\em Cosmological} and {\em Newton} constants $\Lambda$ and $G$, and to the freedom 
	in the choice of the time/space asymmetry.
	Triangulations are then sampled according to 
	a distribution $\propto e^{-S_E}$, with the caveat that 
	$k_4$ is usually traded for a target volume $V$, by 
	adding to $S_E$ a volume fixing term 
	(see Ref.~\cite{cdt_report} for more details).

	In this context, a rich phase structure has been 
	found~\cite{cdt_report,cdt_secondord,cdt_newhightrans,cdt_charnewphase,cdt_toroidalphasediag},
	characterized by four different phases, 
	named respectively $A$, $B$, $C_{dS}$ ({\em de Sitter}) 
	and $C_{b}$ ({\em bifurcation}).
	In the $B$ phase, 
	$V_S$ is concentrated almost in a single slice, 
	while both the $C_{dS}$ and the $C_{b}$ phase
	are characterized by a more regular spatial volume distribution, 
	localized in a so-called ``blob'' with a finite 
	time extension;
	finally, phase $A$ configurations are 
	characterized by multiple and uncorrelated peaks in the spatial volume per slice time $V_S(t)$.
	The bifurcation phase is further differentiated from the $C_{dS}$ phase
	by the presence of two different
	classes of slices which alternate each other 
in {Euclidean time}~\cite{cdt_transfer_matrix,cdt_charnewphase,cdt_newhightrans}.
	
	The transition lines separating the different phases 
	are the candidate places where to search for a continuum limit,
	especially if they are second order. A candidate transition line,
	in this respect, is the one separating the $C_b$ from
	the $C_{dS}$ phase. However, one of the major problems of the 
	CDT program is to find suitable order parameters capable
	of capturing the essential geometrical features around the transition.
	Progress in this direction 
	has been achieved 
	by the study of diffusion
	processes on the triangulations~\cite{cdt_spectdim,cdt_as_from_dimred},
	leading to relevant information such as the 
	{\em spectral dimension} of the triangulations. A
	generalization along this direction has been
	proposed in Ref.~\cite{lbcdt}, consisting in the 
	analysis of the spectrum of the Laplace-Beltrami (LB) 
	operator computed on the triangulations.
	
	The analysis of Ref.~\cite{lbcdt}, limited to the 
	LB operator defined on spatial slices, has shown that
	the various phases can be characterized
	by the presence ($B$) or absence ($A$, $C_{dS}$) of a gap in the spectrum,
	while
	the $C_b$ phase shows 
	the alternance of spatial slices of both types, gapped and 
	non-gapped, which for this reason
	can be named $B$-type and $dS$-type slices.
	The presence of a gap indicates that spatial slices
	are characterized by a high
	connectivity and can be interpreted geometrically as a Universe 
	with an infinite dimensionality at large scales,
	whose diameter grows at most logarithmically as $V_S \to \infty$. 
	On the contrary, the closing of the gap
	can be interpreted as the emergence of an extended Universe with a 
	standard finite dimensionality at large scales.
	In the $C_b$ phase, the alternating slices seem to 
	share {similar} geometries up to some finite length, 
	then differentiating at larger scales. Moreover, 
	the value of the gap seems to change continuously moving
	from the $B$ to the $C_b$ phase, then approaching zero
  towards the $C_{dS}$ phase.
	
	The findings reported above, and the fact that the gap 
	in the spectrum of the LB operator is actually a physical
	quantity with mass dimension two (hence an inverse squared length),
	suggest that it can be used as an order parameter to better
	investigate the $C_b -C_{dS}$ transition and,
	in case it is second order, to characterize the 
	critical behavior around it.
	Having this in mind, the purpose of this study is to put
	the strategy of Ref.~\cite{lbcdt} on a more quantitative
	level, studying the infinite volume limit of the gap and its
	behavior as the transition is approached. Moreover, we will show
	that one can actually find several length scales, all 
  showing a similar {scaling}.

	\emph{Numerical setup} -
	We have investigated CDT with p.b.c. in the Euclidean time 
	direction and an $S^3$ topology for spatial slices. 
	Configurations have been sampled proportionally to $\exp(-S_E)$
	by means of a Metropolis-Hastings algorithm, 
	consisting in a set of local moves (see Ref.~\cite{cdt_report} for
	more details). $N_t = 80$ total space-time slices
	have been taken in all simulations, and the total spatial volume 
	$V_{S,tot} = N_{41}/2$ has been fixed by adding
	to $S_E$ a term $\Delta S = \epsilon (N_{41}-\overline{N}_{41})^2$,
	with $\epsilon = 0.005$, then selecting only 
	configurations with exactly the given target volume 
	$\bar N_{41}/2$.
	
	We have considered the spatial slices of those configurations,
	consisting of sets of glued tetrahedra,
	and computed the eigenvalues of the LB
	operator discretized on them. As in Ref.~\cite{lbcdt}, the
	discretization consists of a linear operator $L$ acting
	on real functions defined on the vertices of the dual graph
	associated with the triangulation. Since 
	any tetrahedron 
	is adjacent to exactly $4$ 
	neighboring tetrahedra, dual graphs are $4$-regular 
	(each vertex is connected with other $4$ vertices), so that 
	$L$ can be written as $L= 4 \cdot \,\mathbbm{1} - A$
	where $A$ is the so-called 
	adjacency matrix, having non-zero unit elements only between 
	pairs of connected vertices.
	In practice, because of this simple form, it suffices to compute 
	the eigenvalues of $A$, which is a sparse matrix, and this
	has been done by means of the 
	`Armadillo' C++ library~\cite{armadillo} with Lapack, Arpack and SuperLU 
	support.

  The smallest eigenvalue is always $\lambda_0 = 0$, and 
  corresponds to a uniform eigenfunction.
  Then, $\lambda_1$ defines the gap
	of the spectrum. In the following we will study 
	$\lambda_1$ and a few other lowest lying eigenvalues
	as a function of the spatial volume $V_S$, trying to extrapolate
	the $V_S \to \infty$ (thermodynamical) limit for each of them. For 
	a regular, extended geometry one expects 
	$\lambda_n \to 0$ in the thermodynamical limit for any
	finite $n$, in particular $\lambda_n \propto 1/D^2$  
	where $D$ is {\em diameter} of the graph
	(maximum over all pairs of vertices of the minimum path length
	connecting the pair).

	We have performed sets of simulations at fixed
	$k_0$ and different values of $\Delta$, chosen so as 
	to stay in the $C_b$ phase but approaching the $C_{dS}$ phase;
	moreover, different values of 
	$\bar N_{41}$ have been considered, to explore the impact on results
	of the total spatial volume. 
	A few simulations in the $B$ phase or in the $C_{dS}$ phase have also 
        been performed, to have a comparison for the infinite volume 
	behavior of the 
	lowest lying eigenvalues in those cases.
	
\begin{figure}[t]
		\includegraphics[width=0.9\columnwidth, clip]{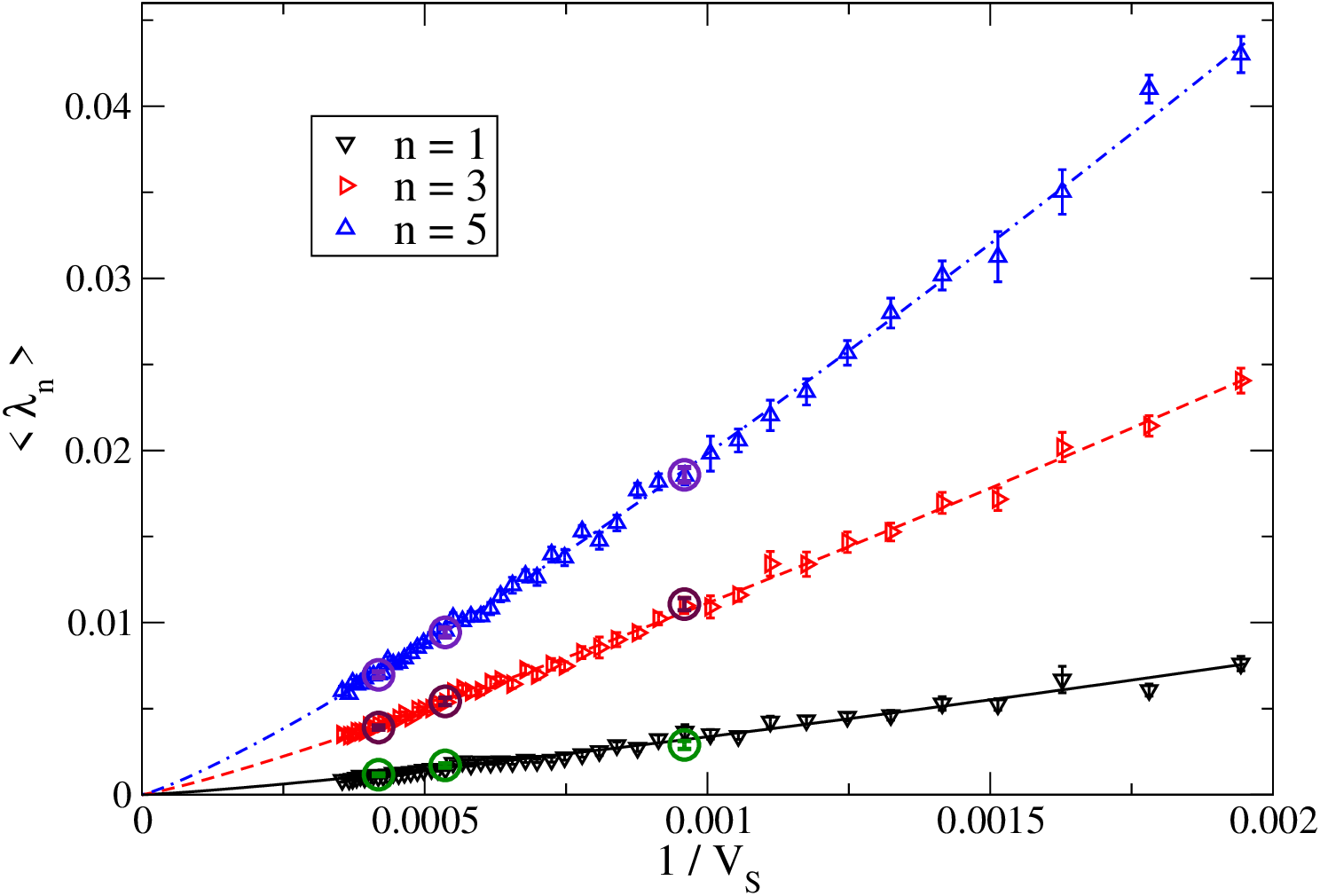}
		\caption{Average eigenvalues of the LB operator
                 on spatial slices as a function 
			of $1 / V_S$ in phase $C_{ds}$ 
 ($k_0 = 0.75$, $\Delta = 0.7$) for $V_{S,tot} = 40$K. 
 We report some data also for $V_{S,tot}~=~60$K (encircled
  points) and best fits 
  to Eq.~(\ref{fit_cds}).
		}
		\label{fig:1}
	\end{figure}

	\emph{Results} -
	We start by analyzing the thermodynamical limit of the
	lowest lying part of the LB spectrum in the $C_{dS}$ phase.
	To that purpose, we have considered a simulation
	performed for $k_0 = 0.75$, $\Delta = 0.7$, where
	the total spatial volume has been fixed to $V_{S,tot} = 4\ 10^4$ (40K).
	Since for each configuration
	most of the spatial volume is distributed over many connected
	slices forming the so-called blob, the volume $V_S$
	of single spatial slice is regularly distributed
	over a wide range going up to a few thousands tetrahedra.
  Because of our finite sample, consisting of about $500$ configurations, 
	this range has been divided in regular bins of volumes, so as to have sufficiently
	populated subsamples in each bin. Average values
	$\langle \lambda_n \rangle$ have then been computed over each bin,
	and results for $n = 1,3,5$ are reported in Fig.~\ref{fig:1},
	statistical errors have been computed by properly
	taking into account autocorrelations
	among triangulations in the sample. For a few bins, we have reported
	also results obtained by fixing a different value of $V_{S,tot}$
        (60K),
	to check that this has no impact on the study
	of the spatial thermodynamical limit.
	
	In the figure we also report best fits according to a power law
	behavior
	\beq
	\langle \lambda_n \rangle = A_n\, V_S^{-2/d_{EFF}} \, ,
        \label{fit_cds}
	\eeq
	all yielding reasonable values of the reduced
	$\chi^2/{\rm d.o.f.}$ and values of $d_{EFF} \simeq 1.6$,
	in agreement with the spectral effective dimension of
	spatial slices at large scales measured in previous studies~\cite{cdt_gorlich,lbcdt};
	similar results are obtained for $n$ up to a few tens.
  This confirms that, in the
	$C_{dS}$ phase, the gap of the LB operator closes
	in the thermodynamical limit, with a
	scaling compatible with the effective spectral dimension
	at large distances.

\begin{figure}[t]
		\includegraphics[width=0.9\columnwidth, clip]{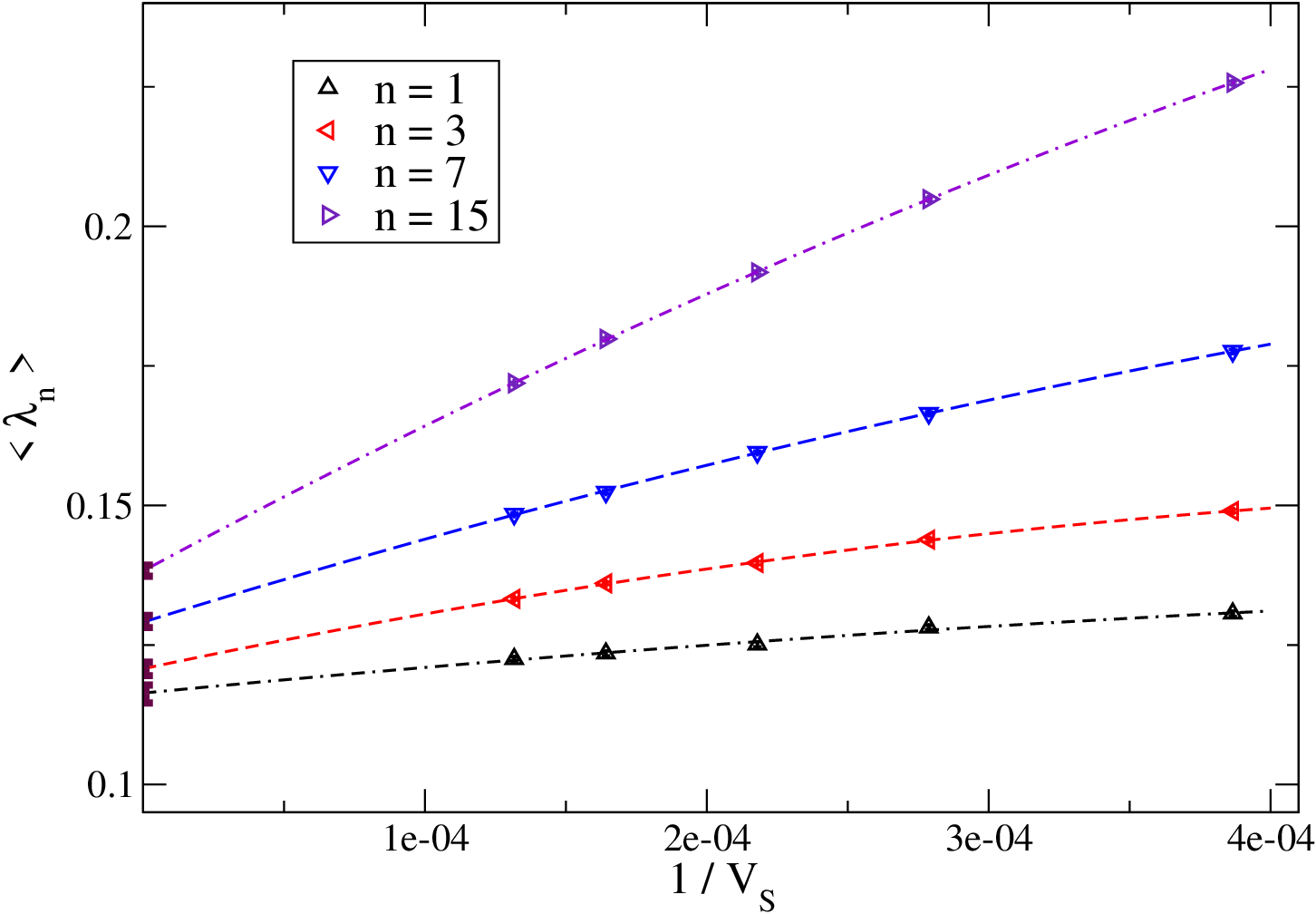}
		\caption{Average eigenvalues as a function 
			of $1 / V_S$ in phase $B$ 
 ($k_0 = 1.0$, $\Delta = -0.2$). Data points at $1 / V_S = 0$ are the 
results of the extrapolation according 
  to Eq.~(\ref{fit_b}).
		}
		\label{fig:2}
	\end{figure}

The situation is quite different in phase $B$.
In this case, most of the total spatial volume
is found in a single slice, so that, in order
to study the thermodynamical limit of the spectrum,
we had to perform
simulations at different values
of $V_{S,tot} = $ 3K, 4K, 5K, 6.5K, 8K. Results obtained for
$\langle \lambda_n \rangle$ on this single slice 
are reported, for a few values of $n$, in Fig.~\ref{fig:2},
$V_S$ in this case being the average volume of the single 
maximal slice:
statistical errors are reported but are not appreciable.
In this case, a smooth extrapolation to the infinite 
volume limit is obtained by allowing simple power corrections
in $V_S^{-1}$:
\beq
\langle \lambda_n \rangle_{V_S} = 
\langle \lambda_n \rangle_\infty + \frac{a_n}{V_S} + \frac{b_n}{V_S^2}
\label{fit_b}
\eeq
and $\chi^2/{\rm d.o.f.} \simeq 1$ is obtained only allowing for 
$b_n \neq 0$.
As already expected from the results of Ref.~\cite{lbcdt}, 
the extrapolated values $\langle \lambda_n \rangle_\infty$ are 
non-zero, as shown in Fig.~\ref{fig:2}. What is more interesting 
is that the extrapolated values for different values
of $n$ do not coincide, i.e.~in the thermodynamical limit
the spectrum above the gap seems to be discrete,
defining a whole hierarchy of length scales for the geometry of the $B$ phase.
Further evidence for this comes from the 
behavior of the volume-normalized spectral density, 
which becomes smaller and smaller as $V_S \to \infty$ 
in the region above the gap, as expected for a discrete spectrum.

\begin{figure}[t]
		\includegraphics[width=0.9\columnwidth, clip]{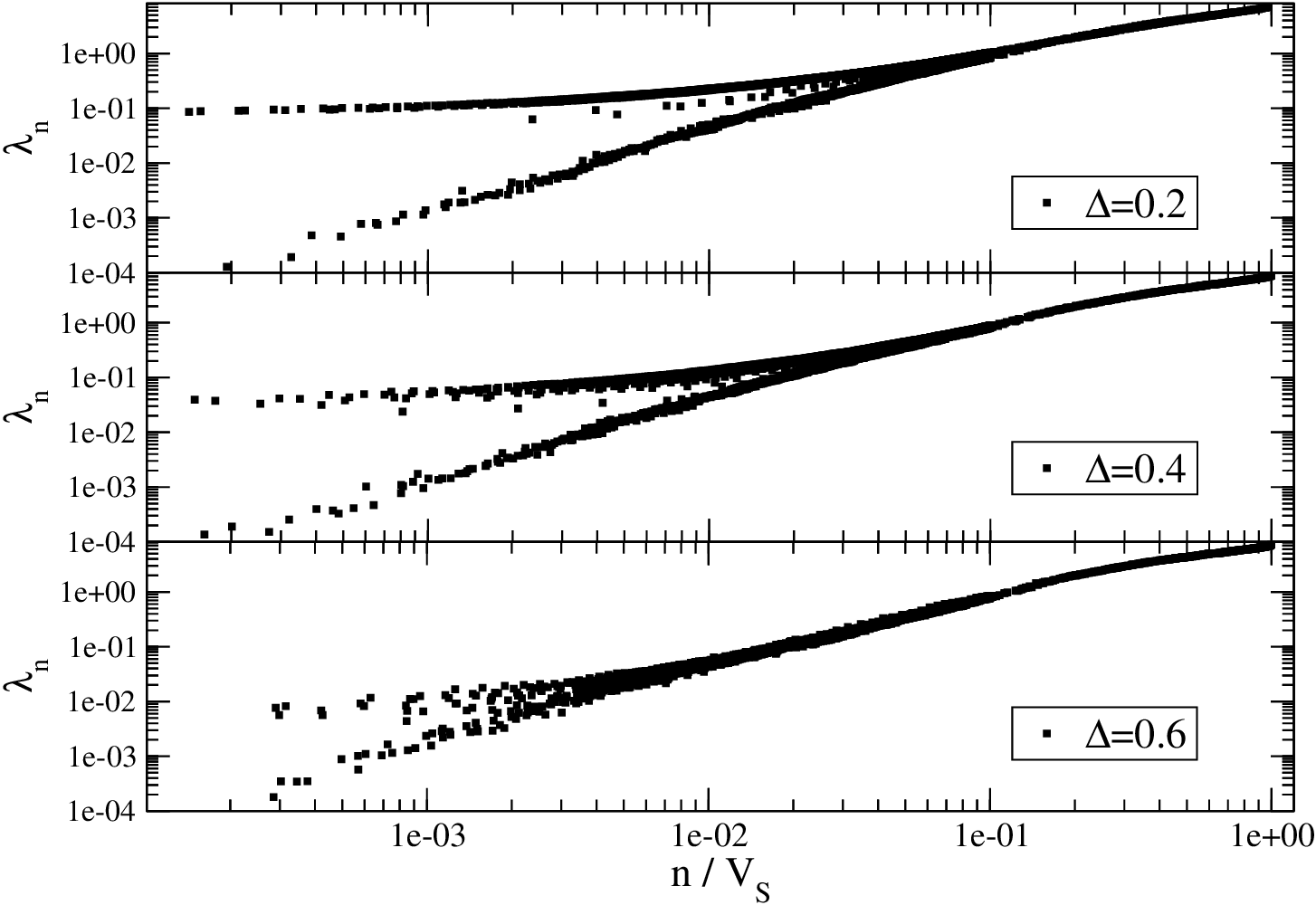}
		\caption{Scatter plot of $\lambda_n$ versus $n/V_S$ for the slices (with spatial volume $V_S > 200$) of single configurations sampled at $k_0=0.75$, and for three different values of $\Delta$ in the $C_b$ phase.	}
		\label{fig:lam_koV_compr}
	\end{figure}

	Our interest is however mostly focused on the 
        intermediate phase $C_b$.
{As discussed above, the two classes of alternating slices
differ mostly at large scales. This is} 
     better enlightened by looking at the scaling profiles 
      where the eigenvalues $\lambda_n$ 
  of the LB operator are reported as a function of the scaling variable 
$n / V_S$: 
  as discussed in Ref.~\cite{lbcdt}, such profiles provide information
   about the effective dimensionality of the spatial triangulations 
  at different scales (smaller values of $n/V_S$ correspond to 
larger scales),  $2/d_{EFF} = {d \log \lambda_n}/{d \log (n/V_S)}$.
In particular, the development of a gap in the LB
operator in the $V_S \to \infty$ limit corresponds to an infinite 
dimensionality at large scales, induced by a high connectivity of the dual
graph.
   In Fig.~\ref{fig:lam_koV_compr} we report the profiles obtained 
  in the $C_b$ phase at $k_0 = 0.75$ and
       for three different values of $\Delta = 0.2, 0.4$ and 0.6:
   the point where two different profiles emerge moves to smaller
  and smaller values of $n/V_S$ as $\Delta$ moves towards the transition 
  to the $C_{dS}$ phase, where the separation in two classes disappears.

\begin{figure}[t]
		\includegraphics[width=0.9\columnwidth, clip]{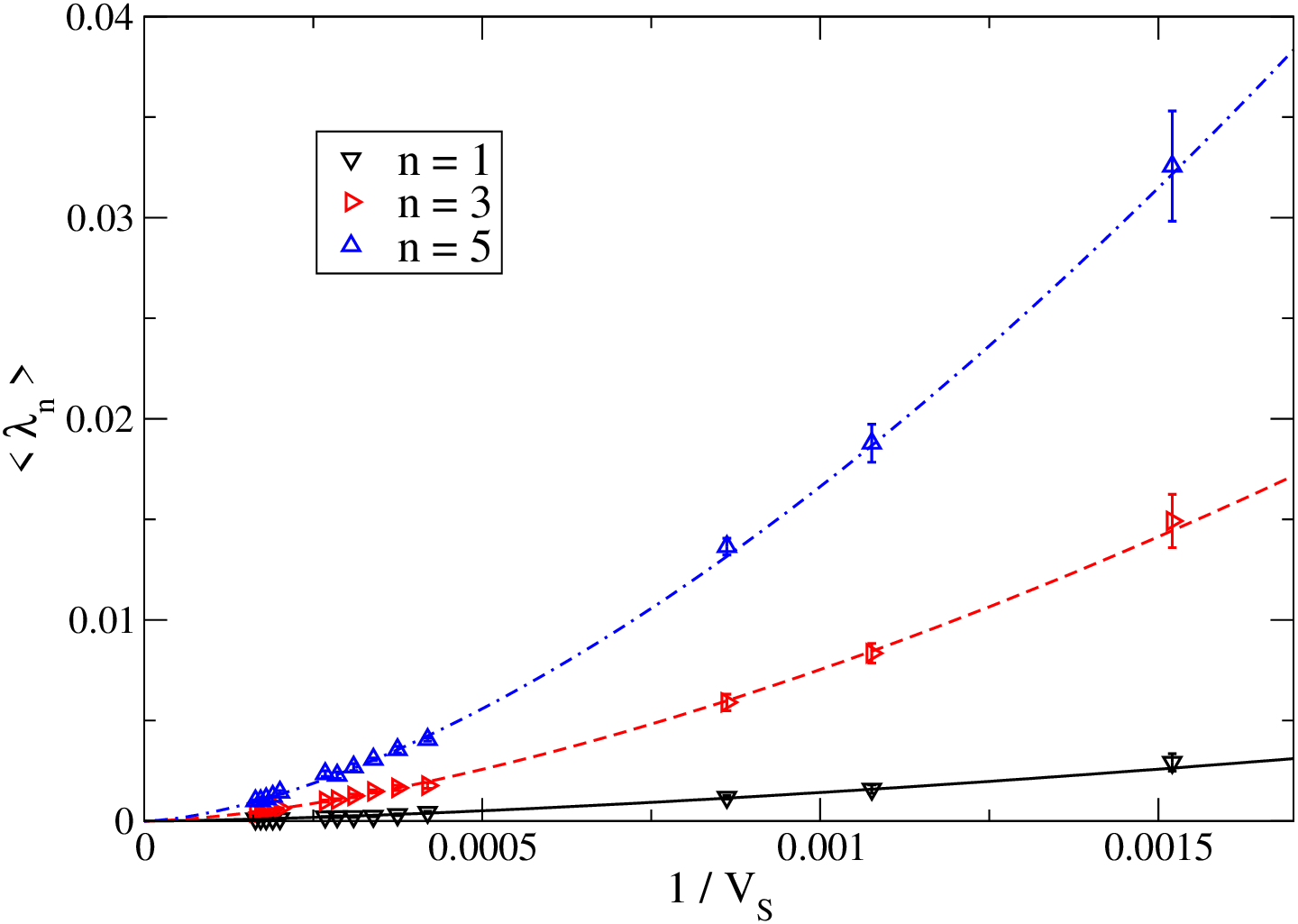}
		\caption{$\langle \lambda_n \rangle (V_S)$ for 
$dS$-like slices in $C_b$ phase ($k_0 = 0.75$, 
$\Delta = 0.4$, $V_{S,tot} = $ 40K).  
 We report also best fits 
  to Eq.~(\ref{fit_cds}).
		}
		\label{fig:3}
	\end{figure}

For each of the lowest lying eigenvalues we have computed 
$\langle \lambda_n \rangle (V_S)$ separately for the two classes of slices, 
starting from $V_S$
large enough to make the separation unambiguous as in Fig.~\ref{fig:lam_koV_compr}. As an example, in Fig.~\ref{fig:3} we report results obtained
for $k_0 = 0.75$ and $\Delta = 0.4$ for {$dS$-like slices: the} 
$V_S \to \infty$ extrapolation can be performed according
to Eq.~(\ref{fit_cds}) in all cases, with $\chi^2/{\rm d.o.f.} \simeq 1$,
confirming the absence of a gap.

\begin{figure}[t]
		\includegraphics[width=0.9\columnwidth, clip]{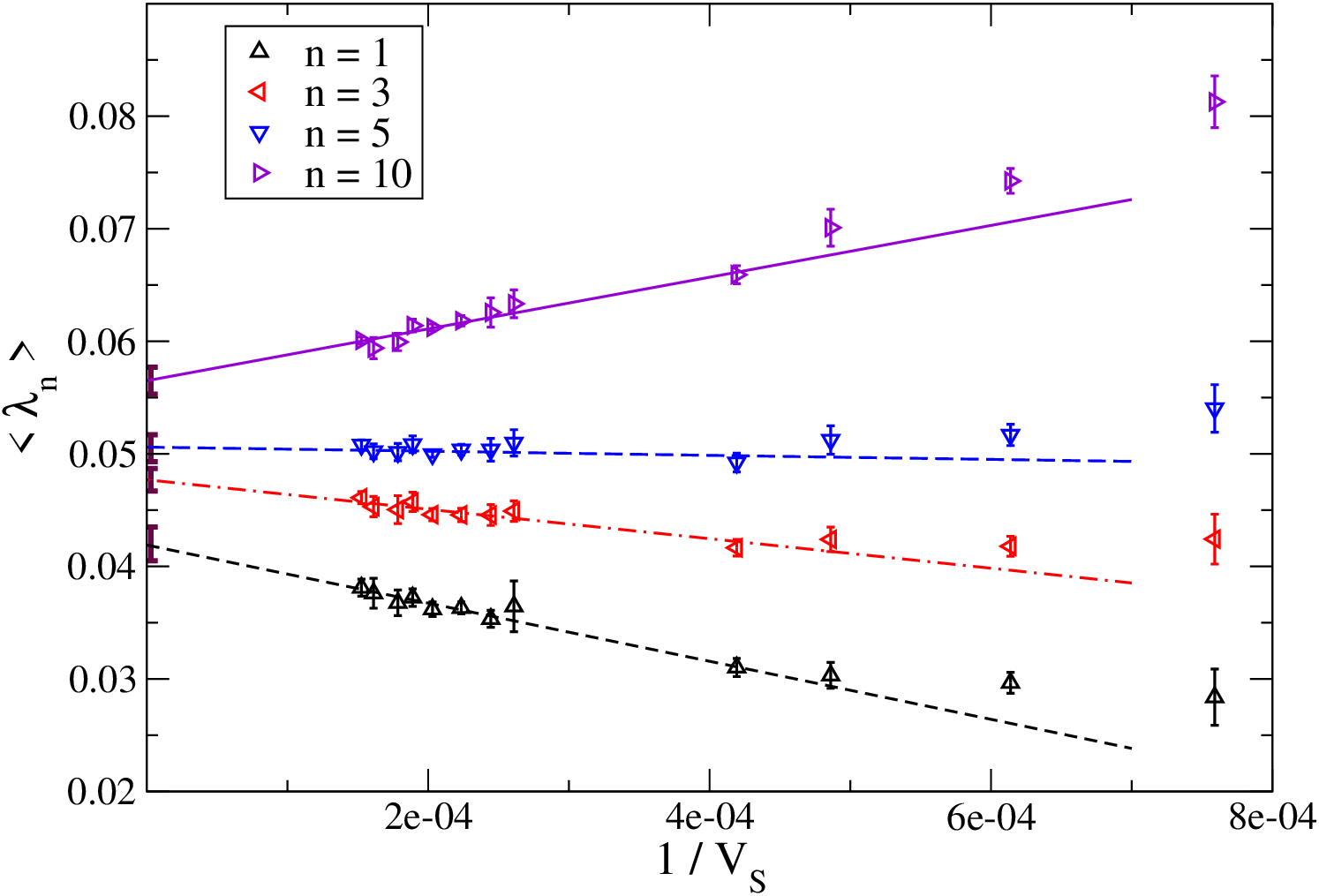}
		\caption{
As in Fig.~\ref{fig:3}, for $B$-like slices.
We report also some $V_S \to \infty$ extrapolations 
with $1/V_S$ corrections (see text).
		}
		\label{fig:4}
	\end{figure}

On the contrary, results for $B$-like slices, which
are reported for the same set of parameters in Fig.~\ref{fig:4},
clearly point to a non-zero $V_S \to \infty$ extrapolation, 
$\langle \lambda_n \rangle_\infty~\neq~0$. Extrapolated values reported in the 
figure have been obtained fitting data according to 
Eq.~(\ref{fit_b}), the reported errors include the systematic 
uncertainty related to the change of the fitted range or the 
inclusion or exclusion of the $1/V_S^2$ term. Even taking these
systematics into account, one notices that the
$\langle \lambda_n \rangle_\infty$ values 
for different $n$ are not compatible,
confirming that also for $B$-like slices 
the lowest lying part of the spectrum 
is likely discrete even in the thermodynamical limit, as for the $B$ phase.

\begin{figure}[t]
		\includegraphics[width=0.9\columnwidth, clip]{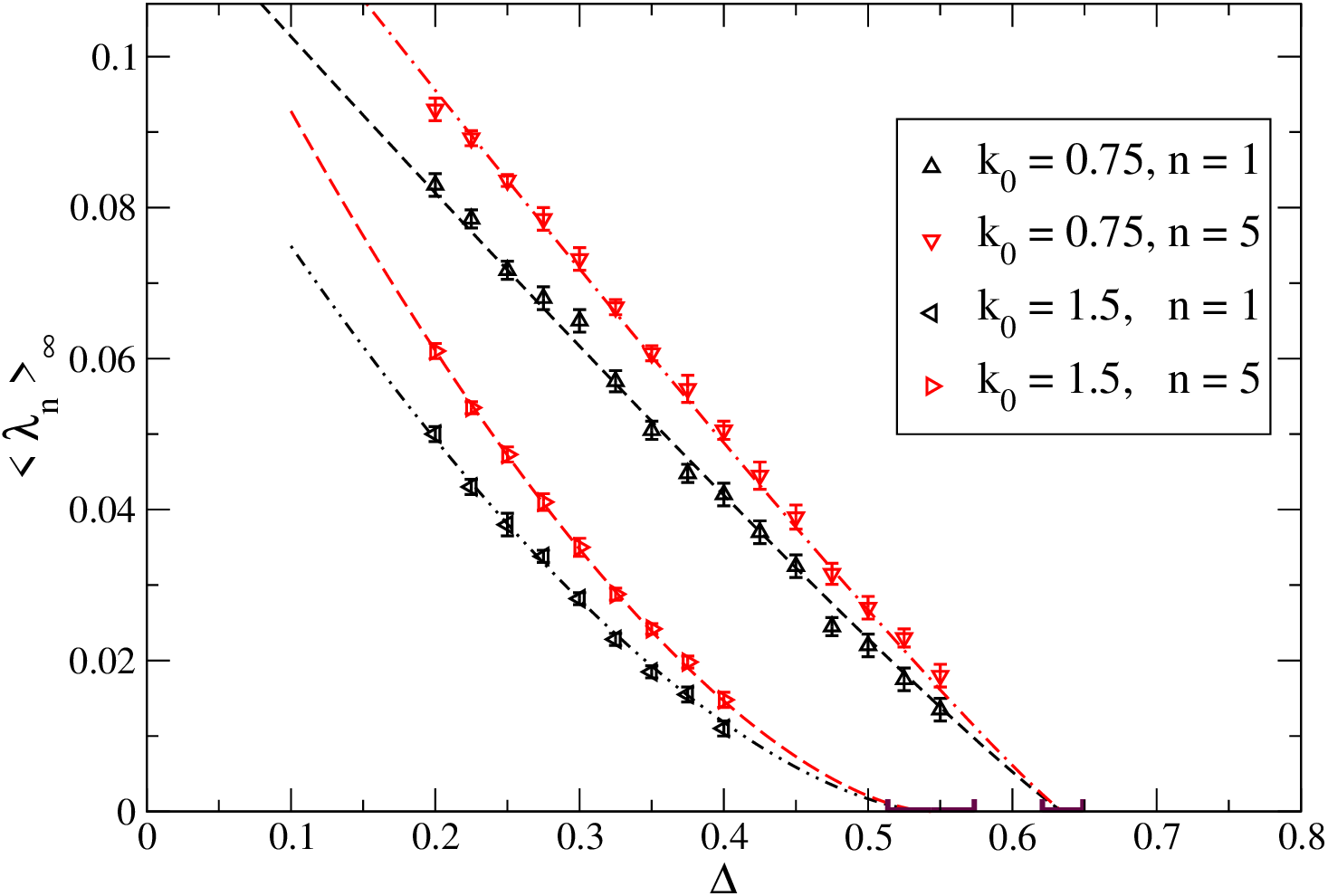}
		\caption{{$\langle \lambda_n \rangle_\infty$ 
 in $B$-like slices as a function of $\Delta$ for different
values of $k_0$ and $n$, together with best fits to Eq.~(\ref{crit_fit}).} 
		}
		\label{fig:6}
	\end{figure}

The main point of our investigation is to adopt 
the non-zero $\langle \lambda_n \rangle_\infty$ of 
$B$-like slices as order parameters approaching zero
at the $C_b - C_{dS}$ transition and probing the 
scaling properties associated with a possible critical point
there. To that purpose, in Fig.~\ref{fig:6} we report
the values obtained for some $\langle \lambda_n \rangle_\infty$ 
($n = 1$ and 5)
as a function of $\Delta$ along two different lines
($k_0 = 0.75$ and $k_0 = 1.5$), crossing the 
$C_b - C_{dS}$ transition line in different points.
On dimensional grounds, different $\langle \lambda_n \rangle_\infty$ 
correspond to different 
inverse squared lengths, which in 
the presence of a critical point 
should scale according to the same
critical behavior. Based on this expectation,
we have tried a fit according to the scaling formula
\beq
\langle \lambda_n \rangle_\infty  = A_n (\Delta - \Delta_c)^{2 \nu}
\label{crit_fit}
\eeq
where only the $A_n$ coefficients depend on $n$.
A combined fit, including $n = 1$ and $n = 5$, yields
$\Delta_c = 0.635(14)$, $\nu = 0.55(4)$ for 
$k_0 = 0.75$ ($\chi^2/{\rm d.o.f.} = 31/26$), and 
$\Delta_c = 0.544(36)$, $\nu = 0.82(12)$ for 
$k_0 = 1.50$ ($\chi^2/{\rm d.o.f.} = 6/14$).
Similar and consistent results are obtained including 
different values of $n$, or if the eigenvalues
are fitted separately; some tension starts to emerge only
when eigenvalues with $n \gtrsim 10$ are included.  
The values of $\Delta_c$ are consistent with those obtained 
analyzing other parameters, based on counting the coordination number
of triangulations, introduced in Ref.~\cite{cdt_signature_change}.
{Our best fits suggests that the critical index $\nu$ could change along
the transition line, however we stress that a global
fit, in which the index is forced to be the same for both $k_0$, works
equally well, yielding $\nu = 0.59(4)$, 
$\Delta_c (k_0 = 0.75) = 0.656(15)$, $\Delta_c (k_0 = 1.5) = 0.479(10)$
with $\chi^2/{\rm d.o.f.} = 47/41$.}
	
	\emph{Concluding remarks} - We have shown that, 
for $B$-like slices in the $C_b$ phase, {as well as for the $B$ phase},
the thermodynamical 
limit of the lowest part of the spectrum of the LB operator
suggests the existence of a discrete set
of inverse squared lengths.
They cannot be considered as proper correlation lengths,
but rather as finite {length} scales which characterize the geometry
of the slices even when $V_S \to \infty$, and have a dependence 
on the bare parameters which is consistent
with 
a common critical behavior 
as the transition to the de Sitter phase is approached.

Our results still do not confirm that the gap 
vanishes continously.
{Indeed,} our determinations of $\langle \lambda_n \rangle_\infty$ 
stop well before $\Delta_c$, since volumes
where eigenvalues of $B$-like and $dS$-like slices are clearly 
distinguished grow as the $C_{dS}$ phase is approached, so that we could not
reliably extrapolate to $V_S = \infty$ close
enough to the transition.
This happens, for instance, for $\Delta = 0.6$ shown in 
Fig.~\ref{fig:lam_koV_compr}: yet, even in this case it is clear
that a gap still exists and is consistent with the critical 
fit reported in Fig.~\ref{fig:6}. Future studies, employing 
larger values of $V_{S,tot}$, {will better clarify the nature
of the possible critical behavior of CDT; moreover, studies at different
values of $k_0$ will clarify if the critical behavior actually
changes along the transition line.}

	\emph{Acknowledgements}
        Numerical simulations have been performed 
   on the MARCONI machine
at CINECA, based on the agreement between INFN
and CINECA (under project INF18\_npqcd) and
at the IT Center of the Pisa University.
        We thank in particular M.~Davini         
        for his technical support.

\end{document}